\documentclass[a4paper,aps,pra,preprintnumbers]{revtex4}
\usepackage[utf8]{inputenc}
\usepackage[english]{babel}
\usepackage[T1]{fontenc}
\usepackage{amssymb,amsfonts,amsmath,mathtext,enumerate,float,dsfont}
\usepackage{graphics,graphicx,epsfig,epstopdf}
\usepackage{caption}
\usepackage{cmap}
\usepackage{multirow}
\usepackage{indentfirst}
\usepackage[usenames]{color}
\usepackage{amsthm}
\usepackage{xcolor}

\begin{document}
\title{Error Correction Using Squeezed Fock States}

\author{S. B. Korolev, E. N. Bashmakova, T. Yu. Golubeva} 
\affiliation{St.Petersburg State University, Universitetskaya nab. 7/9, St.Petersburg, 199034, Russia}

\begin{abstract}

The paper addresses the construction an error correction code for quantum calculations based on squeezed Fock states. It is shown that the use of squeezed Fock states makes it possible to satisfy the Knill-Laflamme (KL) criteria for bosonic error correction codes. It is shown that the first squeezed Fock state corrects both photon loss and dephasing errors better than higher-order states. A comparison of the proposed protocol with an error correction protocol based on the squeezed Schrodinger's cat states is carried out on the basis of the KL cost function. It is shown that the squeezed first Fock state better protects a channel with photon loss and dephasing.
\end{abstract}
\maketitle
\section{Introduction}

The main goal of quantum informatics and quantum optics is to build universal quantum computing. At the moment, quantum computing is in the so-called NISQ (Noisy Intermediate-Scale Quantum) era \cite{Preskill2018QuantumCI}. This era is characterized by the creation of computing systems with a limited number of logic elements and no error correction procedure. When developing medium-scale computing, the primary concern is reducing errors during computations associated with imperfections of the physical systems and the impossibility to completely isolate them from the environment. However, error suppression is not enough to build a full-scale computing procedure. An error correction is necessary to prevent errors from accumulating as operations are performed. Quantum error correction (QEC) codes are used to detect and correct errors \cite{Livingston2022,PhysRevA.52.R2493, PhysRevLett.77.2585, PhysRevLett.102.120501, Braunstein1998}.



At present, error correction protocols on different quantum states have been proposed. For example, encoding information by GKP (Gottesman-Kitaev-Preskill) states \cite{Vasconcelos:10, PhysRevA.64.012310} allow one to construct a protocol for displacement error correction. However, the generation of states with properties approaching GKP states is an extremely difficult experimental task, although there is some progress in this direction \cite{PhysRevA.73.012325, Eaton_2019}.


Schrodinger cat states \cite{Sychev2017,Buzek1995} are another example of quantum states applied for QEC. These states being a superposition of two coherent states, ($|\alpha\rangle$ and $|-\alpha\rangle$) can protect information from photon loss errors. For effective protection, cat states with amplitude $\alpha = 2$ or higher are required \cite{Ralph_2003,Hastrup_2022}. Currently, several protocols have been proposed to generate states similar to Schrodinger cat states \cite{Sychev2017,Ourjoumtsev_cat2007, Polzik2006, Huang2015, Ulanov2016, Gerrits2010, Takahashi2008, Baeva2022,Podoshvedov_2023,Thekkadath2020}. However, the achievable values in the optical range ($|\alpha| \leq 1.9$) \cite{Sychev2017, Ourjoumtsev_cat2007, Huang2015, Ulanov2016, Gerrits2010, Takahashi2008} are insufficient for error correction protocols to work. In the microwave range, Schrodinger cat states with high fidelity and large $\alpha$ amplitude values can be generated \cite{He2023,PhysRevA.99.022302}. However, in such systems, there are other difficulties related to the always-on Kerr nonlinearity suppression, which prevents individual control of logic states \cite{KirchmairG2013Ooqs}.

From the point of view of QEC codes, the squeezed Schrodinger's cat (SSC) states \cite{Grimsmo2020} are being of interest. Based on SSC states, an error correction code was developed that is capable of simultaneously correcting two types of errors: phase errors and photon loss errors. Unlike the traditional Schrödinger cat states, the SSC states should have a large squeezing degree and a small amplitude $\alpha$ for further incorporation into QEC protocols. Protocols for generating SSC state are proposed and discussed in \cite{Wang2022,Bashmakova_2023,PhysRevLett.114.193602,PhysRevA.106.043721,Ourjoumtsev2007}. However, the question of efficient generation of these quantum states with high fidelity and probability remains open.


Error correction codes based on GKP states and Schrodinger's cat states are examples of bosonic QEC codes \cite{PhysRevA.106.022431,Grimsmo2020,Terhal_2020,Wallraff2004,PhysRevLett.111.120501,PhysRevA.94.042332,PhysRevA.64.012310,PhysRevA.93.012315,Mirrahimi_2014,PhysRevA.59.2631,PhysRevA.75.042316,PhysRevA.70.022317,PhysRevA.97.032346} where information is redundantly encoded in quantum oscillator states. At the heart of each QEC code is the requirement to satisfy the Knill and Laflamme conditions (KL-conditions) \cite{PhysRevA.55.900}. When these conditions are approximately met, the code is known as an approximate quantum error correction code \cite{Schlegel_2022,Reinhold}. All bosonic codes known to date are approximate QEC codes.  

Squeezed Fock (SF) states \cite{Olivares_2005} are an example of well-studied non-Gaussian states. However, such states have not, to the authors' knowledge, been considered yet as a resource for QEC. In our work, we demonstrate that it is possible to construct an approximate error correction code based on SF states. Using the KL cost function \cite{Schlegel_2022,Reinhold} as a measure, we first examine different SF states and answer the question which SF states are better suited for correcting photon loss errors and dephasing errors. Next, using this function, we compare our protocol with a protocol using SSC states.


 \section{Error Correction Using Squeezed Fock States}
Let us say that we have a quantum channel in which there are two types of errors: particle loss and dephasing. The evolution of the state described by the density matrix $\hat{\rho}$ in this channel is given by the following Lindblad master equation \cite{Lindblad1976}:
\begin{align}
\partial_{t} \hat{\rho}(t) & =\mathcal{L} \hat{\rho}(t)=\kappa_{1} \mathcal{D}[\hat{a}] \hat{\rho}(t)+\kappa_{2} \mathcal{D}\left[\hat{a}^{\dagger} \hat{a}\right] \hat{\rho}(t), \label{eq_lin}
\end{align}
where
\begin{align}
\mathcal{D}[\hat{J}] \hat{\rho}(t) & =\hat{J} \hat{\rho}(t) \hat{J}^{\dagger}-\frac{\hat{J}^{\dagger} \hat{J} \hat{\rho}(t)+\hat{\rho}(t) \hat{J}^{\dagger} \hat{J}}{2},
\end{align}
the annihilation $\hat{a}$ and creation $\hat{a}^{\dagger}$ operators obey the canonical commutation relation $\left[\hat{a},\hat{a}^{\dagger}\right]=1$. The superoperators $\mathcal{D}[\hat{a}]$ and $\mathcal{D}\left[\hat{a}^{\dagger} \hat{a}\right]$ describe the particle loss and dephasing at rates of $\kappa_1$ and $\kappa_2$, respectively. The solution to equation (\ref{eq_lin}) can be written in the form of the Kraus decomposition \cite{KRAUS1971311}:
\begin{align}
\hat{\rho}(t)=\sum_{j=0}^{\infty} \hat{K}_{j} \hat{\rho}(0) \hat{K}_{j}^{\dagger}.
\end{align}
In the case when the error accumulation rate is small $\kappa_{1,2} t \ll 1$, only three operators can be left in the decomposition:
\begin{align}
&\hat{K}_{0}=\hat{I}-\frac{\kappa_{1} t}{2} \hat{a}^{\dagger} \hat{a}-\frac{\kappa_{2} t}{2}\left(\hat{a}^{\dagger} \hat{a}\right)^{2}, \\
&\hat{K}_{1}=\sqrt{\kappa_{1} t} \hat{a}, \quad \hat{K}_{2}=\sqrt{\kappa_{2} t} \hat{a}^{\dagger} \hat{a} .
\end{align}
It turns out that at low error rates ($\kappa_{1,2} t \ll 1$), in a channel with particle loss and dephasing, it is possible to protect information if one is able to correct the following set of elementary errors:
\begin{align} \label{Error_set}
    \mathcal{E}=\left\{\hat{I}, \hat{a}, \hat{a}^{\dagger} \hat{a},\left(\hat{a}^{\dagger} \hat{a}\right)^{2}\right\}.
\end{align}
Thus, our goal is to encode information in a way that protects it from these errors. To this end, we consider a bosonic QEC code \cite{PhysRevA.106.022431,Grimsmo2020,Terhal_2020,Wallraff2004,PhysRevLett.111.120501,PhysRevA.94.042332,PhysRevA.64.012310,PhysRevA.93.012315,Mirrahimi_2014,PhysRevA.59.2631,PhysRevA.75.042316,PhysRevA.70.022317,PhysRevA.97.032346}, in which we encode two logical states of a qubit into a quantum oscillator. In the bosonic QEC we consider, the code space to be a two-dimensional subspace of an infinite-dimensional Hilbert space. The code space is characterized by two basis states (code words) that encode the logical states $|0\rangle_L$ and $|1\rangle_L$. When individual errors act on the code words, they move to other states from the error subspace. The error subspace vectors associated with different errors must be orthogonal to each other and to the code subspace vectors. All these requirements are conveniently written in the form of KL conditions \cite{PhysRevA.55.900}:
\begin{align}
\left\langle i_{L}\left|\hat{E}_{a}^{\dagger} \hat{E}_{b}\right| j_{L}\right\rangle =\delta_{i j} \alpha_{a b},
\label{KL}
\end{align}
where $ i,j\in \lbrace 0,1\rbrace$,  and $\hat{E}_{a}, \hat{E}_{b}\in \mathcal{E}$, and $ \alpha_{a b}$ is a matrix that does not depend on $i$ and $j$. The presented condition is a necessary and sufficient condition for the recovery of the code words after the action of the errors.

To correct for both particle loss and dephasing errors, we use bosonic QEC code based on SF states. These states are defined as follows:
\begin{align}
    |r,n\rangle=\hat{S}(r)|n\rangle,
\end{align}
where $\hat{S}(r)$ is the squeezing operator, $r$ is the squeezing parameter, and $|n\rangle$ is the Fock state. To correct for photon loss and dephasing errors, we require states with specific parity \cite{PhysRevLett.111.120501,PhysRevA.94.042332} and structure in phase space \cite{PhysRevA.106.022431,Leviant2022quantumcapacity}. SF states satisfy all these properties.


As code words, we will consider two states of the form:
\begin{align}
    &|0_{L};n\rangle=\hat{S}\left(r\right)|n\rangle,\\
    &|1_{L};n\rangle=\hat{S}\left(-r\right)|n\rangle,
\end{align}
The represented states are squeezed in orthogonal directions on the phase plane.

The code words we entered depend on two parameters: the squeezing degree and the Fock state number. These two parameters can be used to optimize the error correction protocol. Below we will study the dependence of code words on the number of the Fock state, but now let us look at the KL conditions for the first SF state.

\section{Quantum error correction code based on the first squeezed Fock state}
Let us consider correcting both particle loss and dephasing errors using the first SF state. In this case, the code words are the following states:
\begin{align} 
    &|0_{L};1\rangle=\hat{S}\left(r\right)|1\rangle \label{FFS_1},\\
    &|1_{L};1\rangle=\hat{S}\left(-r\right)|1\rangle. \label{FFS_2}
\end{align}
Using these states, we can write KL conditions for the orthogonality of the states after errors from set (\ref{Error_set}) act on them. All these conditions are presented in Table \ref{tab:KL_First_Fock}.
\begin{table}[h!]
    \centering
    \begin{tabular}{|c|c|c|c|c|}
    \hline
       $ \hat{E}_a \textbackslash \hat{E} _b$ & $\hat{I}$  & $\hat{a}$ & $\hat{a}^{\dag}\hat{a}$ & $\left(\hat{a}^{\dag}\hat{a}\right)^2$\\
        & & & &  \\
       \hline
       $\hat{I}$ &  $\displaystyle\frac{1}{\cosh ^{{3}/{2}} 2r}$ & 0 & $\displaystyle\frac{3-\cosh 2 r}{2 \cosh^{{5}/{2}}2 r}$ & $\displaystyle\frac{25-12 \cosh 2 r-5\cosh 4 r}{8 \cosh^{{7}/{2}}2 r}$  \\
        & & & &  \\
        \hline
        $\hat{a}$ & 0 & $\displaystyle\frac{3-\cosh 2 r}{2 \cosh^{{5}/{2}}2 r}$ & 0 & 0  \\
        & & & &  \\
        \hline
       $\hat{a}^{\dag}\hat{a}$  & $\displaystyle\frac{3-\cosh 2 r}{2 \cosh^{{5}/{2}} 2r}$ & 0 & $\displaystyle\frac{25-12 \cosh 2 r-5\cosh 4 r}{8 \cosh^{{7}/{2}}2 r} $ & $\displaystyle\frac{282-129 \cosh 2 r-138 \cosh 4 r+17 \cosh 6 r}{32 \cosh^{{9}/{2}}2 r}$ \\
       & & & &  \\
       \hline
       $\left(\hat{a}^{\dag}\hat{a}\right)^2$  & $\displaystyle\frac{25-12 \cosh (2 r)-5\cosh 4 r}{8 \cosh^{{7}/{2}}2 r} $ & 0 & $\frac{282-129 \cosh 2 r-138 \cosh 4 r+17 \cosh 6 r}{32 \cosh^{{9}/{2}}2 r}$ & $\frac{4203-1560 \cosh 2 r-3236 \cosh 4 r+600 \cosh 6 r+121 \cosh 8 r}{128 \cosh^{{11}/{2}}2 r} $\\
       & & & &  \\
       \hline
    \end{tabular}
    \caption{KL conditions $\left\langle 0_{L};1\left|\hat{E}_{a}^{\dagger} \hat{E}_{b}\right| 1_{L};1\right\rangle$ and $\left\langle 1_{L};1\left|\hat{E}_{a}^{\dagger} \hat{E}_{b}\right| 0_{L};1\right\rangle$ for code words based on the first SF state.}
    \label{tab:KL_First_Fock}
\end{table}
\noindent The KL condition associated with the error norm is always satisfied for code words (\ref{FFS_1}) and (\ref{FFS_2}). I. e., for any pair of errors from set (\ref{Error_set}), the following statement is true:
\begin{align}
    \left\langle 0_{L};1\left|\hat{E}_{a}^{\dagger} \hat{E}_{b}\right| 0_{L};1\right\rangle=\left\langle 1_{L};1\left|\hat{E}_{a}^{\dagger} \hat{E}_{b}\right| 1_{L};1\right\rangle.
\end{align}
The exact fulfillment of this condition is an undoubted advantage of this code. This distinguishes our proposed approximate code from others in which this condition is approximately satisfied \cite{Grimsmo2020,PhysRevX.10.011058,Hastrup_2022}. It is important to note that the states used for encoding can be accurately generated experimentally \cite{SFock}. This means that the exact equality of error norms condition is not violated in the experiment.

\begin{figure}[H]
    \centering
    \includegraphics{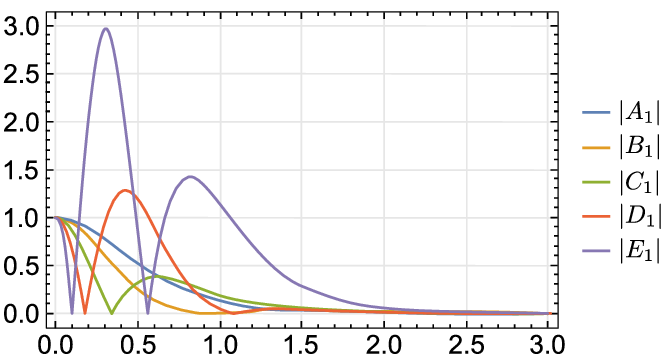}
    \caption{Dependence of the absolute value of KL conditions for the orthogonality on the squeezing parameter $r$.}
    \label{fig:A_1-E_1}
\end{figure}

For code words based on the first squeezed Fock state to be useful for correcting particle loss and dephasing errors, all elements in Table \ref{tab:KL_First_Fock} must be small.  I.e., we need to require the following expressions to tend to zero:
\begin{align}
    &A_1=\left\langle i_{L};1| j_{L};1\right\rangle=\frac{1}{\cosh ^{{3}/{2}} 2r},\\
    &B_1=\left\langle i_{L};1\left|\hat{a}^{\dagger} \hat{a}\right| j_{L};1\right\rangle=\frac{3-\cosh 2 r}{2 \cosh^{{5}/{2}}2 r},\\
    &C_1=\left\langle i_{L};1\left|\left(\hat{a}^{\dagger} \hat{a}\right)^2\right| j_{L};1\right\rangle=\frac{25-12 \cosh 2 r-5\cosh 4 r}{8 \cosh^{{7}/{2}}2 r}, \\
    &D_1=\left\langle i_{L};1\left|\left(\hat{a}^{\dagger} \hat{a}\right)^3\right| j_{L};1\right\rangle=\frac{282-129 \cosh 2 r-138 \cosh 4 r+17 \cosh 6 r}{32 \cosh^{{9}/{2}}2 r},\\
    &E_1=\left\langle i_{L};1\left|\left(\hat{a}^{\dagger} \hat{a}\right)^4\right| j_{L};1\right\rangle=\frac{4203-1560 \cosh 2 r-3236 \cosh 4 r+600 \cosh 6 r+121 \cosh 8 r}{128 \cosh^{{11}/{2}}2 r},
\end{align}
where $i\neq j=0,1$. For the convenience of analyzing the presented functions, let us build their absolute value in Fig. \ref{fig:A_1-E_1}.
The graph shows that different functions have different zeros. At the same time, the zeros do not coincide, and the zero of one function corresponds to large values of the other. In addition, function $\left|A_1\right|$ has no zeros, indicating the absence of orthogonality of code words for any squeezing parameter $r$. Thus, the only way to achieve orthogonality of states after the action of errors is to use states with a large parameter $r$. For example, for $r=2$, we get the following values: $A_1\approx 7 \cdot 10^{-3}$, $B_1\approx -3 \cdot 10^{-3}$, $C_1 \approx -9 \cdot 10^{-3}$, $D_1 \approx 10^{-2}$, $E_1 \approx 5 \cdot 10^{-2}$. We see that they are small but not equal to zero. This means that the code we have presented is an approximate bosonic QEC   \cite{PhysRevA.56.2567,PhysRevA.81.062342,PhysRevLett.104.120501,PhysRevLett.107.080501,PhysRevA.86.012335}. All bosonic QEC codes belong to this type of code \cite{PhysRevA.106.022431,Grimsmo2020,Terhal_2020,Wallraff2004,PhysRevLett.111.120501,PhysRevA.94.042332,PhysRevA.64.012310,PhysRevA.93.012315,Mirrahimi_2014,PhysRevA.59.2631,PhysRevA.75.042316,PhysRevA.70.022317,PhysRevA.97.032346}. 


Thus, the proposed QEC code based on the first SF state can correct low-rate particle loss and dephasing errors under large squeezing parameters.

\section{Comparison of quantum error correction codes based on squeezed Fock states with different numbers}
In the previous section we considered first SF states as code words. Now, let us look at other SF states and understand which ones are best suited for correcting particle loss and dephasing errors. Appendix \ref{appen_A} presents KL conditions for encodings using different SF states (with different $n$). From the presented conditions, it is clear that for any $n$, KL conditions on the error norm are exactly satisfied, while the KL conditions on error orthogonality are approximately satisfied for large $r$. All this means that codes based on the SF state with an arbitrary $n$ can be used to correct photon loss and dephasing errors.

Once we have established that SF states of arbitrary number n are suitable for QEC, we can proceed to compare them. As a measure for comparing codes, we will use the KL cost function proposed in \cite{Reinhold} and used to compare two codes in the paper \cite{Schlegel_2022}.

\begin{figure}[H]
    \centering
    \includegraphics[scale=0.85]{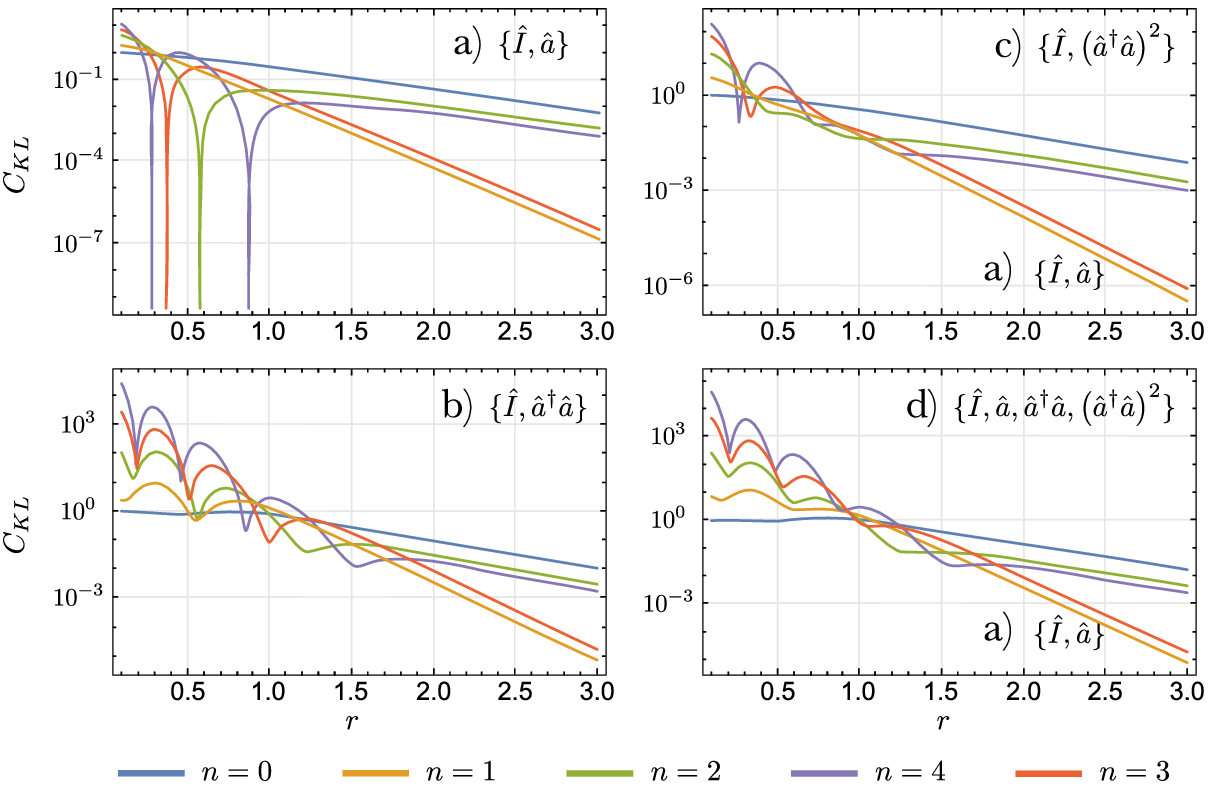}
    \caption{KL cost functions depending on the squeezind parameter for various SF states. The figure shows the KL cost function for a different set of errors:  a) $\lbrace \hat{I},\hat{a}\rbrace$, b) $\lbrace \hat{I},\hat{a}^\dag \hat{a}\rbrace$, c) $\lbrace \hat{I},\left(\hat{a}^\dag \hat{a}\right)^2\rbrace$, d) $\lbrace \hat{I}, \hat{a}, \hat{a}^\dag \hat{a}, \left(\hat{a}^\dag \hat{a}\right)^2\rbrace$.}
    \label{fig:CKLn}
\end{figure}

The KL cost function is given by the following expression:
\begin{align}
    C_{KL} \left(\left\lbrace \hat{E}\right\rbrace\right)=\sum _{a,b}\left| f_{00ab}-f_{11ab}
\right|^2+\left| f_{01ab}\right|^2,
\end{align}
where the KL tensor is given by the expression
\begin{align}
    f_{ijab}=\left\langle i_L \left|\hat{E}^\dag _a\hat{E}_b\right|j_L\right\rangle,
\end{align}
and $\lbrace\hat{E}\rbrace \subset \mathcal{E}$ is represented by a set of error operators over which the sum is taken. I. e., we can evaluate codes by their ability to correct both individual types of errors and their totality. In the limit of perfect code, for which the KL conditions (\ref{KL}) are exactly satisfied, the value of the KL cost function is equal to zero. The larger the KL cost function value, the worse the approximate code corrects errors \cite{Schlegel_2022,Reinhold}. Fig. \ref{fig:CKLn} shows the dependencies of the KL cost function for various sets of errors and different code words depending on the squeezing parameter.
%

Fig. \ref{fig:CKLn}a shows that to correct particle loss errors, we can achieve zeros at specific finite values of the parameter $r$ for $n=2,3,4$. In other words, when used to encode the second, third, fourth, and higher orders of SF states, we have an ideal code that protects information from the loss of particles. However, in this case, we can not protect information from dephasing. From Fig. \ref{fig:CKLn}b and \ref{fig:CKLn}c, it is clear that in the region $r\in \left[0.2,1\right]$, codes with $n=2,3,4$ have values close to the maximum. Given that our goal is to correct particle loss and dephasing errors, we need to pay attention to Fig. \ref{fig:CKLn}d. It shows the KL cost functions for both types of errors. It is clear from this figure that the KL cost function takes minimum values for the first SF state starting from a certain squeezing parameter $r>1.7$. This means that for protecting information from two types of errors, the first SF state with a high squeezing degree is better suited (among all the cases considered).

The statement we have obtained can be strengthened by saying that the first SF state best (among SF states with different numbers) protects information from loss and dephasing errors. This statement follows directly from two facts. First, as was shown in Appendix \ref{appen_A}, the KL conditions for SF states of odd numbers tend to zero faster than the same conditions for even numbers. Second, it follows from Table \ref{tab:KL_2n+1_Fock} that the first SF state has the smallest asymptote for the KL cost function (for large values of $r$) among all possible odd SF states.


Thus, we can conclude that if we have a channel with only particle loss error, then SF states with numbers greater than one will be the best for information protection. If, in the experiment, we manage to generate these states with certain values of $r$ (values corresponding to the zeros of the KL cost function), then we will get an ideal QEC code for the particle loss error. However, since the dips corresponding to the zeros of the cost function (see Fig. \ref{fig:cat_n}a) are very narrow, we will need high accuracy in generating the states. If we have a channel with two types of errors, then it is better to use the first SF state with the maximum possible squeezing degree.


\section{Comparison of error correction protocols based on squeezed Schrodinger's cat states and squeezed Fock states}

Let us compare our protocol with the protocol based on the SSC states. This comparison is motivated by the fact that the protocol using the SSC states is able to correct particle loss and dephasing errors simultaneously \cite{Grimsmo2020,Schlegel_2022} . Furthermore, this code is similar to our proposed code in terms of experimental realization. Both SF states \cite{SFock}  and states close to the SSC states can be obtained experimentally (in an optical scheme with measurements). Thus, we compare two codes that can be realized experimentally.

Two states are used as code words in the protocol based on the SSC states:
\begin{align}
    &|0_{L,SCS}\rangle=\frac{1}{N_{+}}\left(|\alpha,r\rangle+|-\alpha,r\rangle\right),\\
    &|1_{L,SCS}\rangle=\frac{1}{N_{-}}\left(|\alpha,r\rangle-|-\alpha,r\rangle\right),
\end{align}
where $|\alpha,r\rangle=\hat{S}\left(r\right)\hat{D}\left(\alpha\right)|0\rangle$ are squeezed coherent states, and $N_{\pm}=\sqrt{2\left(1\pm e^{-2\alpha ^2 e^{2r}}\right)}$ is the normalization factor. As we found out in the previous section, the best way to correct the two types of errors is to use the encoding with the first SF state. We will compare this encoding with the encoding based on the SSC states.

\begin{figure}[H]
    \centering
    \includegraphics[scale=0.85]{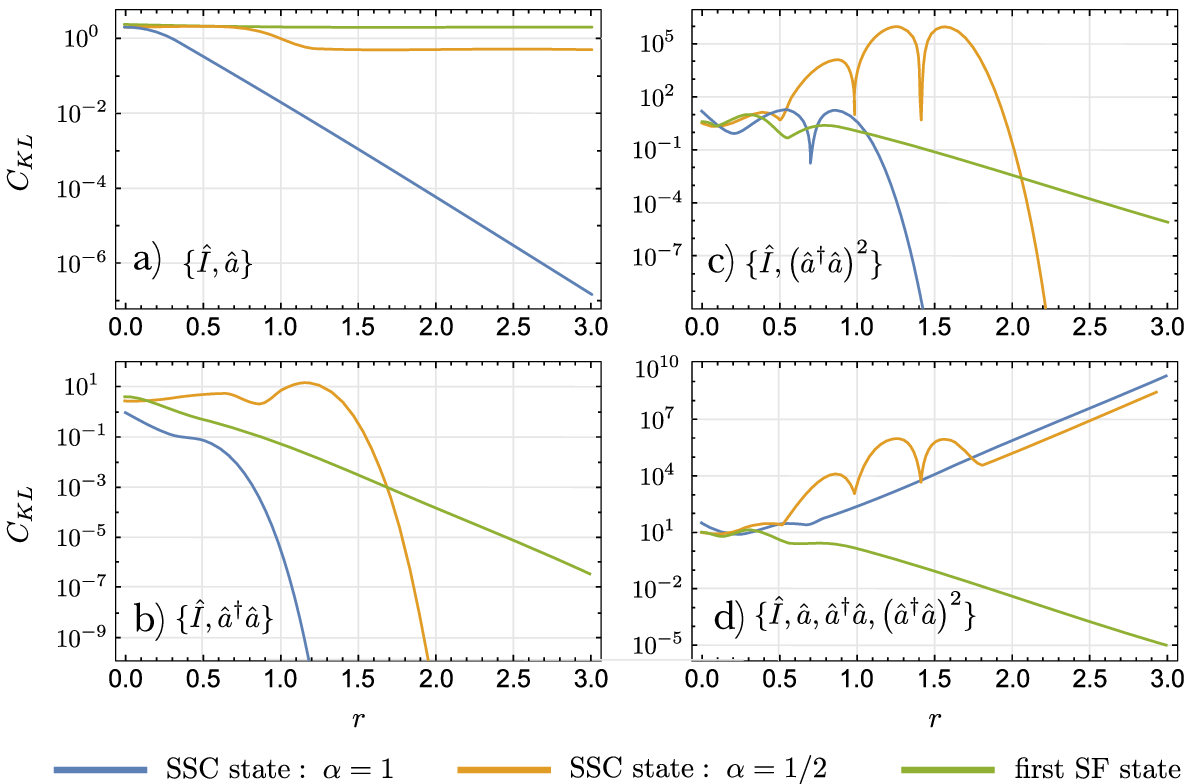}
    \caption{KL cost functions depending on the squeezing parameter for three encodings: two based on the SSC states with different amplitudes $\alpha$, and one based on the first SF state. The figure shows KL cost functions for different sets of errors:  a) $\lbrace \hat{I},\hat{a}\rbrace$, b) $\lbrace \hat{I},\hat{a}^\dag \hat{a}\rbrace$, c) $\lbrace \hat{I},\left(\hat{a}^\dag \hat{a}\right)^2\rbrace$, d) $\lbrace \hat{I}, \hat{a}, \hat{a}^\dag \hat{a}, \left(\hat{a}^\dag \hat{a}\right)^2\rbrace$.}
    \label{fig:cat_n}
\end{figure}

The KL cost function graphs for two different encodings are shown in Fig. \ref{fig:cat_n}.
Fig. \ref{fig:cat_n}a shows that the first SF state corrects particle loss errors significantly better. Indeed, for almost all squeezing parameters $r$, the KL cost function is smaller for the encoding with the first SF state than for the encoding with the SSC states.

As for the dephasing errors (Figs. \ref{fig:cat_n}b and \ref{fig:cat_n}c), it all depends on the relationship between the amplitude and the squeezing parameter of the SSC state. For example, for $\alpha=0.5$ and  $r<1.7$, the first SF state corrects the phase error better (Fig. \ref{fig:cat_n}b). However, for large amplitudes of W, the situation is the opposite. The ability to correct the $\left(\hat{a}^{\dagger}\hat{a}\right)^2$ error also depends on the value of the parameter $\alpha$.  When the amplitude value is large, the code using the SSC states corrects this type of error better compared to the first SF state (Fig. \ref{fig:cat_n} c). At the same time, the larger the amplitude of the cat state, the smaller its ability to correct particle loss errors (see Fig. \ref{fig:cat_n}a). The combination of these two factors leads to the fact that the SF state being generally better at correcting errors from the $\mathcal{E}$ set (see Fig. \ref{fig:cat_n}d).

It is important to note that it does not follow from Fig. \ref{fig:cat_n}d that the code based on the SSC state is bad at correcting errors of both types. To talk about the quality of the code, we should perform a more detailed analysis with estimation of the channel fidelity and construction of recovery operations. The results obtained by using the KL cost function only allow us to compare encodings. Thus, the obtained estimates allow us to declare that, other things being equal, the encoding with the first SF state corrects errors better than the encoding with the SSC states.

\section{Conclusion}
The paper addressed the problem of quantum error correction (QEC) in a quantum channel with particle loss and dephasing errors are present. We have shown that squeezed Fock (SF) states can be used to encode information in such a channel. These states have a certain parity and a structure in the phase space. That is why we considered them as the main resource for QEC. 


We have shown that the approximate Knill-Laflamme (KL) conditions are satisfied for the SF states with arbitrary numbers $n$. The conditions on the orthogonality of the errors are satisfied when the squeezing of SF states tend to infinity. The condition on the norm of the errors for SF states is always satisfied. 


To compare different quantum codes with each other, we exploit the KL cost function. This is a function indicating how much the KL condition is violated for selected code words. Using this function, one can give a quantitative measure for evaluating different code words.


Applying this measure, we have shown that the first SF state is the best for information protection in a channel with both particle loss and dephasing errors. In this case, the squeezing degree should be large enough. In this paper, we have found that for a squeezing parameter $r>1.7$, the code based on the first SF state performs better than the code based on SF states with any other number.


Considering a channel with only particle loss errors, we obtain a perfect QEC code using SF states with a certain squeezing degree and number $n>1$. For this code, the KL conditions are perfectly satisfied. However, it is important to note that implementing such a code in an experiment is quite challenging since we have to tune the parameter $r$ precisely.


We compared code words based on the first SF state with code words based on the squeezed Schrodinger's cat states. We have shown that the code based on the first SF state is better suited for information protection in a channel where both particle loss and dephasing errors are present. We demonstrated that for the same squeezing degree of the two states, the KL cost function of the SF state is smaller for a channel with two types of errors. In other words, the first SF states better protect the information in a channel where both particle loss and dephasing errors are present.
\\
\\

\section*{Funding}
This work was financially supported by the Russian Science Foundation (Grant No. 24-22-00004).

\section*{Acknowledgments}

The authors are grateful to Prof. A. K. Tagantsev for fruitful discussion and valuable advice.

\section*{Disclosures}
The authors declare no conflicts of interest.

\section*{Data availability} Data underlying the results presented in this paper are not publicly available at this time but may be obtained from the authors upon reasonable request.

\bibliography{nongaussian}  

\appendix
\section{KL conditions for encoding using squeezed Fock states with arbitrary n} \label{appen_A}
Let us see what the KL conditions (\ref{KL}) look like for an encoding using  SF states with an arbitrary number $n$. 
\begin{align}
    &|0_{L};n\rangle=\hat{S}\left(r\right)|n\rangle,\\
    &|1_{L};n\rangle=\hat{S}\left(-r\right)|n\rangle,
\end{align}
As for the first SF state, the error norm condition is exactly satisfied for any $n$:
\begin{align}
    \left\langle 0_{L};n\left|\hat{E}_{a}^{\dagger} \hat{E}_{b}\right| 0_{L};n\right\rangle=\left\langle 1_{L};n\left|\hat{E}_{a}^{\dagger} \hat{E}_{b}\right| 1_{L};n\right\rangle
\end{align}
It turns out that the KL conditions for orthogonality depend on the parity of the codes used. The KL conditions for error orthogonality for encoding using odd SF states are presented in Table \ref{tab:KL_2n+1_Fock}. 
\begin{table} [H]
    \centering
    \begin{tabular}{|c|c|c|c|c|}
    \hline
       $ \hat{E}_a \textbackslash \hat{E} _b$ & $\hat{I}$  & $\hat{a}$ & $\hat{a}^{\dag}\hat{a}$ & $\left(\hat{a}^{\dag}\hat{a}\right)^2$  \\
       \hline
        $\hat{I}$ &  $\frac{\Gamma \left(n+1\right)}{\Gamma \left(\frac{n+1}{2}\right)^2}\frac{4\sqrt{2} e^{-3 r}}{ 2^n}$ & 0 & $\frac{\Gamma \left(\frac{n}{2}\right)}{\Gamma \left(\frac{n+1}{2}\right)}\frac{\sqrt{2} n e^{-3 r}}{\sqrt{\pi} }$ & $\frac{\Gamma \left(\frac{n+2}{2}\right)}{\Gamma \left(\frac{n+1}{2}\right)}\frac{5 \sqrt{2}   e^{-3 r}}{\sqrt{\pi}}$  \\
        \hline
        $\hat{a}$ & 0 & $\frac{\Gamma \left(\frac{n}{2}\right)}{\Gamma \left(\frac{n+1}{2}\right)}\frac{\sqrt{2} n e^{-3 r}}{\sqrt{\pi} }$ & 0 & 0  \\
        \hline
       $\hat{a}^{\dag}\hat{a}$  & $\frac{\Gamma \left(\frac{n}{2}\right)}{\Gamma \left(\frac{n+1}{2}\right)}\frac{\sqrt{2} n e^{-3 r}}{\sqrt{\pi} }$ & 0 & $\frac{\Gamma \left(\frac{n+2}{2}\right)}{\Gamma \left(\frac{n+1}{2}\right)}\frac{5 \sqrt{2}   e^{-3 r}}{\sqrt{\pi}}$ & $\frac{\Gamma \left(\frac{n}{2}+1\right)^2}{  \Gamma (n+1)}\frac{2^{n} 17  e^{-3 r}}{\pi \sqrt{2}
       }$ \\
       \hline
       $\left(\hat{a}^{\dag}\hat{a}\right)^2$  & $\frac{\Gamma \left(\frac{n+2}{2}\right)}{\Gamma \left(\frac{n+1}{2}\right)}\frac{5 \sqrt{2}   e^{-3 r}}{\sqrt{\pi}}$ & 0 & $\frac{\Gamma \left(\frac{n}{2}+1\right)^2}{  \Gamma (n+1)}\frac{2^{n} 17  e^{-3 r}}{\pi \sqrt{2}
       }$ & $\frac{\Gamma \left(\frac{n}{2}+1\right)^2}{\Gamma (n+1)} \frac{121\ 2^{n}  e^{-3 r}}{2\sqrt{2}\pi}$\\
       \hline
    \end{tabular}
    \caption{Asymptotic behavior of KL conditions $\left\langle 0_{L};2n+1\left|\hat{E}_{a}^{\dagger} \hat{E}_{b}\right| 1_{L};2n+1\right\rangle$ and $\left\langle 1_{L};2n+1\left|\hat{E}_{a}^{\dagger} \hat{E}_{b}\right| 0_{L};2n+1\right\rangle$ for large $r$ for code words based on SF states with odd numbers.}
    \label{tab:KL_2n+1_Fock}
\end{table}
We see that all non-zero conditions tend to zero for large squeezing parameter $r$, as $\mathcal{O}\left(e^{-3r}\right)$. 

 The KL conditions for error orthogonality for encoding using even SF states are presented in Table \ref{tab:KL_2n_Fock}. 
\begin{table} [H]
    \centering
    \begin{tabular}{|c|c|c|c|c|}
    \hline
       $ \hat{E}_a \textbackslash \hat{E} _b$ & $\hat{I}$  & $\hat{a}$ & $\hat{a}^{\dag}\hat{a}$ & $\left(\hat{a}^{\dag}\hat{a}\right)^2$  \\
       \hline
        $\hat{I}$ &  $\frac{\sqrt{2} \Gamma \left(\frac{n+1}{2}\right)}{\sqrt{\pi}\Gamma \left(\frac{n+2}{2}\right)} e^{-r}$ & 0 & $\frac{\Gamma \left(\frac{n+1}{2}\right)}{\sqrt{2 \pi } \Gamma \left(\frac{n+2}{2}\right)}e^{-r}$ & $\frac{\Gamma \left(\frac{n+1}{2}\right)}{\Gamma \left(\frac{n+2}{2}\right)} \frac{e^{-r}}{2\sqrt{2 \pi}}$  \\
        \hline
        $\hat{a}$ & 0 & $\frac{\Gamma \left(\frac{n+1}{2}\right)}{\sqrt{2 \pi } \Gamma \left(\frac{n+2}{2}\right)}e^{-r}$ & 0 & 0  \\
        \hline
       $\hat{a}^{\dag}\hat{a}$  & $\frac{\Gamma \left(\frac{n+1}{2}\right)}{\sqrt{2 \pi } \Gamma \left(\frac{n+2}{2}\right)}e^{-r}$ & 0 & $\frac{\Gamma \left(\frac{n+1}{2}\right)}{\Gamma \left(\frac{n+2}{2}\right)} \frac{e^{-r}}{2\sqrt{2 \pi}}$ & $\frac{\Gamma \left(\frac{n+1}{2}\right)}{\Gamma \left(\frac{n+2}{2}\right)} \frac{5 e^{-r}}{4 \sqrt{2 \pi }}$ \\
       \hline
       $\left(\hat{a}^{\dag}\hat{a}\right)^2$  & $\frac{\Gamma \left(\frac{n+1}{2}\right)}{\Gamma \left(\frac{n+2}{2}\right)} \frac{e^{-r}}{2\sqrt{2 \pi}}$ & 0 & $\frac{\Gamma \left(\frac{n+1}{2}\right)}{\Gamma \left(\frac{n+2}{2}\right)} \frac{5 e^{-r}}{4 \sqrt{2 \pi }}$ & $\frac{\Gamma \left(\frac{n+1}{2}\right)^2}{  \Gamma (n+1)}\frac{ 2^{n}  17 e^{-r}}{8 \pi \sqrt{2}}$ \\
       \hline
    \end{tabular}
    \caption{Asymptotic behavior of KL conditions $\left\langle 0_{L};2n\left|\hat{E}_{a}^{\dagger} \hat{E}_{b}\right| 1_{L};2n\right\rangle$ and $\left\langle 1_{L};2n\left|\hat{E}_{a}^{\dagger} \hat{E}_{b}\right| 0_{L};2n\right\rangle$ for large $r$ for code words based on SF states with even numbers.}
    \label{tab:KL_2n_Fock}
\end{table}
From the table we see that all non-zero conditions tend to zero at large squeezing parameter $r$, as $\mathcal{O}\left(e^{-r}\right)$. Comparing Table \ref{tab:KL_2n+1_Fock} with Table \ref{tab:KL_2n_Fock}, we can conclude that the use of SF states with odd numbers is better because the KL conditions for them tend to zero faster than the same conditions for even numbers SF states.

Furthermore, it is not difficult to show that every non-zero element of Table \ref{tab:KL_2n+1_Fock} takes its minimum value in the case where $n=1$. This means that the KL conditions are best satisfied for the first SF state. It also follows that for large $r$ the KL cost function for the first SF state will have the minimum value among all possible SF states. 

%
%

\end{document}